\documentclass[runningheads,a4paper]{llncs}

\usepackage{amssymb}
\usepackage{graphicx}

\usepackage{amsmath}

\usepackage{syntax}
\usepackage{listings}
\usepackage{color, colortbl}

\definecolor{codegreen}{rgb}{0,0.6,0}
\definecolor{codegray}{rgb}{0.5,0.5,0.5}
\definecolor{codepurple}{rgb}{0.58,0,0.82}
\definecolor{backcolour}{rgb}{0.95,0.95,0.92}

\lstdefinestyle{mystyle}{
  backgroundcolor=\color{backcolour},
  commentstyle=\color{codegreen},
  numberstyle=\tiny\color{codegray},
  stringstyle=\color{codepurple},
  basicstyle=\ttfamily,
  breakatwhitespace=false,         
  breaklines=true,                 
  captionpos=b,                    
  keepspaces=true,                 
  numbersep=5pt,                  
  showspaces=false,                
  showstringspaces=false,
  showtabs=false,                  
  tabsize=2
}
\lstdefinelanguage{Eel}
{morekeywords={Log, If, PIf, For, PFor, Else, While, Unroll, LogPush, Def, Return, GotoIfEq, GotoIfNeq, Goto, PGotoIfEq, PGotoIfNeq, PGoto, Label, PLabel, Call, ADD, SUB, MULT, LPUSH, LPOP, PUSH, POP, NEG, XOR, SWAP, MOVE, AND, OR, GOTO, GOTOIF, GOTOIFN, CMFRM, CMFRMIF, CMFRMIFN},
sensitive=true,
morecomment=[l]{//},
morecomment=[s]{/*}{*/},
morestring=[s]{`}{'}
}
\def\rowheading#1{\multicolumn{2}{l}{\cellcolor{black}\textcolor{white}{\textbf{#1}}}}
\def\columnheading#1{\multicolumn{1}{c}{\cellcolor{blue}\textcolor{white}{\textbf{#1}}}}
\def\unheading#1{\rm #1}

\usepackage{url}
\urldef{\mailsa}\path|{ntyagi, jaysonl, edemaine}}@mit.edu|

\begin{document}

\mainmatter

\title{Toward an Energy Efficient Language and Compiler for (Partially) Reversible Algorithms}

\titlerunning{Energy Efficient Language and Compiler}

\author{Nirvan Tyagi%
\thanks{MIT Computer Science and Artificial Intelligence Laboratory,
      32 Vassar Street, Cambridge, MA 02139, USA,
      \protect\url{{edemaine,jaysonl,geronm,ntyagi}@mit.edu}.
      Supported in part by the MIT Energy Initiative and by
      MADALGO --- Center for Massive Data Algorithmics ---
      a Center of the Danish National Research Foundation.}
\and Jayson Lynch\footnotemark[1] 
\and Erik D. Demaine\footnotemark[1]}
\authorrunning{Energy Efficient Language and Compiler}

\institute{MIT CSAIL}

\toctitle{Reversible Computation 2016}
\tocauthor{Authors' Instructions}
\maketitle

\begin{abstract}
We introduce a new programming language for expressing reversibility, Energy-Efficient Language (Eel), geared toward algorithm design and implementation. Eel is the first language to take advantage of a partially reversible computation model, where programs can be composed of both reversible and irreversible operations. In this model, irreversible operations cost \emph{energy} for every bit of information created or destroyed. To handle programs of varying degrees of reversibility, Eel supports a log stack to automatically trade energy costs for space costs, and introduces many powerful control logic operators including protected conditional, general conditional, protected loops, and general loops.  In this paper, we present the design and compiler for the three language levels of Eel along with an interpreter to simulate and annotate incurred energy costs of a program.
\end{abstract}

\section{Introduction}

Continued progress in technology has created a world where we are increasingly dependent on computers and computing power. Computer use is greatly increasing and thus becoming a significant energy expenditure for the world. It is estimated that computing consumes more than 3\% of the global electricity consumption \cite{Pavel}, growing at a steady rate. Improved energy efficiency of computers translates to savings in money and environmental toll. Additionally, improved energy efficiency would lead to increased longevity of batteries or use of a smaller battery for the same lifespan. This applies most directly to portable devices such as laptops, mobile phones, and watches where battery size and life are of the utmost importance. Finally, improved energy efficiency would lead to faster CPUs. The main bottleneck in increasing clock speeds are cooling constraints. With decreased energy consumption, we can expect to be able to increase CPU speed by roughly the same factor with the same cooling. Given these many motivations, continued improvement of the energy efficiency of computation is an important research field.

\textbf{Fundamental limits to efficiency.}
If computer energy efficiency continues to progress at a similar rate, we will expect to hit a fundamental limit based in physics and information theory known as Landauer's limit \cite{Lan61} within the next 15-60 years. Landauer gives a lower limit for the energy cost of losing one bit of information of $kT\ln 2$ units of energy where $k$ is Boltzmann's constant and $T$ is temperature: at room temperature $T=20^\circ$C, approximately $2.8\cdot 10^{-21}$ joules or $7.8\cdot 10^{28}$ kilowatt hours. Our current computation systems depend on computing models that require the erasure of information (boolean circuits, random access machines). 

\textbf{Reversible computation model.}
Reversible computation, where the inputs can be recovered from the outputs and no bits of information are lost, is the common approach studied in order to improve computing efficiency beyond Landauer's limit. In this paper, we consider a variant of the traditional reversible computation model we call \emph{partially} reversible computation \cite{us}, allowing for both reversible and irreversible operations. Traditional models of computation include two main constraints in the asymptotic analysis of algorithms, time and space. However, with the introduction of partially reversible computation, a new natural metric emerges, which we call \emph{energy}. In this model, from Landauer's principle, reversible computation is free, but creating or destroying bits of information costs energy. The energy cost of an operation is equal to each bit of information created or destroyed and comes from the change in information entropy from inputs to outputs.

\textbf{Energy-efficient language (Eel).}
We break down the results into two main parts. First, we present a new reversible programming language, Eel. Eel is composed of three language levels with the high-level based on Python and the low-level based on PISA \cite{Vieri99,Vieri98}. Eel is the first programming language to take advantage of partially reversible computation. Past research on reversible programming languages has focused on computation which is performed fully reversibly. Eel allows operations to erase bits and incur energy cost. Eel also allows users to indicate operations for reversal and will automatically store the proper information in a log stack (separate from the stack). In addition, we introduce a number of high level control logic operators of varying degrees of reversibility. With the partial reversibility model, Eel brings the time, space, and energy tradeoffs to the forefront.

Second, we present a compiler and interpreter in Java for Eel. We describe the compilation techniques used between the Eel language levels to handle the high level control logic. We also describe the interpreter technique to simulate and annotate the energy costs of program execution. Since general purpose fully-reversible computers are still years away from development, an interpreter that simulates energy costs is valuable for algorithm development and implementation.

\section{Previous Work} \label{sec:previous-work}
The study of reversible computation to circumvent Landauer's limit has been a broad area of research for a number of years, ranging from development of reversible hardware, analysis of reversible algorithmic theory, to development of reversible programming languages and computer architecture. The origins of the field can be traced back to Lecerf \cite{lecerf63} and to Bennett \cite{Ben73}. Early theory results show that any algorithm can be made reversible with either quadratic space overhead \cite{Ben89} or with exponential time overhead \cite{LMT97,BTV01}. However, it is unknown whether or not any given algorithm can be converted to a reversible version maintaining the same time and space constraints. Some models introduce an algorithmic complexity based on information erased during a computation\cite{LV96,us} laying a foundation for partially reversible computing.

Past research on reversible programming languages has focused on fully reversible programming languages and architectures. The first high-level reversible programming languages developed were Janus \cite{Lutz,Yoko09} and R\cite{Fra99}. We understand that there are a set of properties that must be held by all reversible languages \cite{Yoko08}, and that these properties are satisfied in Janus.
Fully reversible computer architectures have been built. Pendulum \cite{Vieri99,Vieri98}, the first reversible architecture built, was introduced along with a reversible low-level instruction set, PISA, which is used as a basic reversible instruction set in many future works. An improved reversible architecture\cite{Axel07} compatible with PISA introduces a novel technique for handling branches, previously handled with traces, using space to keep track of program counter jumps. Most recently, this architecture has been further improved with the development of Bob \cite{Thom12} using a slightly modified version of PISA known as BobISA, providing more efficient branch handling and address calculation. The Eel low level language uses an instruction set based on PISA expanded to support irreversible operations.

There exist both a reversible self-interpreter for Janus \cite{Yoko07} and a partial evaluator for Janus \cite{Mog11,Mog12}. There also exist general techniques for compilation between reversible languages \cite{Axel11} and compilation of regular programs to reversible programs \cite{Peru99}.

Although Eel is still in its early stages of development, it is designed to provide a unique perspective to reversible programming and, specifically, algorithm development. Where Janus is a powerful and mature language for fully reversible programming, the partial reversibility of Eel opens up a whole new set of options for developers. Eel brings forward the tradeoff for irreversible logic between energy cost and space cost in the log stack. Eel introduces new high level control logic operators that represent different options on the energy-space tradeoff spectrum. Additionally, Eel allows for partial reversals of the program for each code block, a useful feature to have for algorithm development. While this is also possible in Janus, it requires a nesting of function calls and uncalls. Overall, the aim of Eel is to provide a reversible language geared toward algorithm design and implementation in a partially reversible model. 

\section{Language Design}
In this section we discuss some of the design decisions that went into the language. There is an overview of what operations are exposed in each of the three languages written. We also discuss how reversing computation is notated.

\subsection{Logging and Unrolling} \label{sec:log-unroll}

Eel supports partially reversible programs consisting both of logic blocks that will be reversed and logic blocks that will only be executed in the forward direction. In the high level, to denote a section of code to be reversed, it is placed inside of a \texttt{Log} statement to form a log block. The high level is organized into code blocks of varying levels of nesting. An \texttt{Unroll} statement indicates the reverse execution of pending log blocks within the block. All log blocks within a code block must be unrolled before exiting to the previous nesting level. This unroll method can be generalized to allow for a more complex unrolling order. See future works section for further discussion.

Some operations in a log block, such as assignments and branching, are not easily reversible. Eel handles these operations by automatically logging information (storing trace information) about the operation using auxilary space when executed in the forward direction. Upon reversal, the logged information is used to reverse the operation and is then zeroed out. The notion of using auxiliary space, or a ``history" stack (we call \emph{log stack}), to make irreversible computation reversible has been used in the past for irreversible operations such as memory overwrites and switch branching \cite{Zul01,Stod09}. We extend this idea to support higher level control logic operators and see how different assumptions on control logic conditions change what information needs to be logged. A basic example of how the log stack is used for an irreversible assignment operation is in Figure \ref{fig:unrollexample}. The assignment operation is irreversible since the information previously stored at the memory location is overwritten. To make the assignment operation reversible, the previous value is stored in and retrieved from the log stack using \texttt{LPUSH} and \texttt{LPOP} operations. These operations increment and decrement the log pointer and maintain the memory location at the top of the log stack to be zero.

Eel automatically handles logging information for supported control logic operators and irreversible operations, but for more advanced functions additional information may need to be stored. Eel high level provides the \texttt{LogPush} command to push an item onto the log stack. \texttt{LogPush} can be used to make user-defined functions supported reversibly.

\begin{figure}[!htb] \label{fig:unrollexample}
\centering
\begin{minipage}{.5\textwidth}
    
\begin{lstlisting}[language=Eel]
`High Level'
Log:
    x += 1
Unroll

`Low Level'
ADD(x,1)
SUB(x,1)  //Unroll starts


\end{lstlisting}

\end{minipage}%
\begin{minipage}{0.5\textwidth}

\begin{lstlisting}[language=Eel]
`High Level'
Log:
    x = 1
Unroll

`Low Level'
LPUSH(x)
ADD(x,1)
SUB(x,1)  //Unroll starts
LPOP(x)
\end{lstlisting}

\end{minipage}
\caption{Basic example of using log stack and not using log stack for reversal. \texttt{LPUSH} and \texttt{LPOP} perform the appropriate operation to the log stack and zero out the previous location.}
\end{figure}

\subsection{Language Levels} \label{sec:lang-levels}
Eel is designed with three different levels exposing different levels of complexity. The high level language provides a Python-like syntax and common control operators for algorithm development. This is meant to seem familiar and to hide some of the difficulties of working in a partially reversible environment. The intermediate level is stripped down to a simpler set of commands and attempts to resemble working in transdichotomious RAM models of computation. By necessity it also exposes some fairly mechanical parts of execution such as the program and log stacks. It reduces the control logic to a series of jumps. This tries to compromise between readability, clear resource calculations, and expressive power. The low level gives a basic instruction set one might imagine for a semi-reversible computer based off of PISA. Here we have a small number of basic operations where the time, space, and energy costs of each line are clear.

\textbf{High Level.} \label{sec:high-level}
The high level handles the partial reversibility of Eel with the \texttt{Log} and \texttt{Unroll} keywords. Placing operations inside of a \texttt{Log} block indicates to the compiler that these operations will be reversed. If there is an irreversible operation or control logic operator in a \texttt{Log} block, specific information is stored (logged) in the log stack. During an unroll, this information is used to properly reverse the operation and zeroed out.

Variables are not strongly typed and do not have explicit declaration. Instead variables are created the first time they are used. There are interesting questions concerning performance and ease-of-use with respect to typing in reversible programming languages; however, we have not yet been able to explore this substantially.

Basic control logic operators, such as conditionals and loops, are supported at the high level. However, different keywords are used to describe operators of different reversibility. For example, a \emph{protected} conditional is completely reversible and does not require any space in the log, but requires assumptions on the usage of the condition variables. A \emph{general} conditional, with no such assumptions, is not inherently reversible and requires a single bit of information to be stored in the log for reversibility. Table \ref{tab:high-summary} summarizes the operators available at the high level and the space required in the log stack to be made reversible. The reversibility of these high level control logic operators is studied in more detail in a companion paper \cite{us}. In an attempt to simplify the control logic, we note that the current protected operators in the high level provide less expressiveness than other languages such as Janus. However, the intermediate language is fully expressive, and future iterations of the high level can include more complex operators built from the intermediate level. Figure \ref{fig:highgrammar} shows the grammar of the high level.

Basic control logic operators, such as conditionals and loops, are supported at the high level. However, different keywords are used to describe operators of different reversibility. For example, a \emph{protected} conditional is completely reversible and does not require any space in the log, but requires assumptions on the usage of the condition variables. A \emph{general} conditional, with no such assumptions, is not inherently reversible and requires a single bit of information to be stored in the log for reversibility. Table \ref{tab:high-summary} summarizes the operators available at the high level and the space required in the log stack to be made reversible. The reversibility of these high level control logic operators is studied in more detail in a companion paper \cite{us}. In an attempt to simplify the control logic, we note that the current protected operators in the high level provide less expressiveness than other languages such as Janus. However, the intermediate language is fully expressive, and future iterations of the high level can include more complex operators built from the intermediate level. Figure \ref{fig:highgrammar} shows the grammar of the high level.

\begin{figure}[!htb]\label{fig:highgrammar}
\centering

\begin{grammar}

<program> ::= <b> \hfill block

<b> ::= <s>* \hfill statement sequence

<s> ::= $x$ ${\otimes}{=}$ <e> | $x =$ <e> \hfill assignment
\alt `PIf'( <e> ): <b> (`Else': <b>)? \hfill protected conditional
\alt `If'( <e> ): <b> (`Else': <b>)? \hfill general conditional
\alt `PFor'( <s>, <e>, <s> ): <b> \hfill protected for loop
\alt `For'( <s>, <e>, <s> ): <b> \hfill general for loop
\alt `While'( <e> ): <b> \hfill general while loop
\alt `Def' $q(x, \ldots, x)$: <b> \hfill function definition
\alt $q(x, \ldots, x)$ \hfill function call
\alt `Log': <b> \hfill log block
\alt `Unroll' \hfill unroll

<e> ::= $c$ | $x$ | <e> $\odot$ <e> \hfill expression

<$\otimes$> ::= $+$ | $-$ | $*$ \hfill operators

<$\odot$> ::= $\otimes$ | / | $\leq$ | $\geq$ | $\neq$ | $==$
\end{grammar}
\caption{Eel high level grammar, where
  $x \in$ Vars, $q \in$ FxnIds, $c \in$ IntConsts}
\end{figure}

\begin{table}[!htb]
\caption{Summary of high level control keywords and the amount of space in the log stack required to make reversible if appearing in a log block.}
\label{tab:high-summary}
\centering
\begin{tabular}{|l|l|c|c|}
\columnheading{Control Operator} & \columnheading{Keyword} & \columnheading{Log \unheading (bits)} & \columnheading{Sec.}  \\ \hline
Protected Conditional & \texttt{PIf}(cond) & $0$ & \ref{sec:conditionals}\\ \hline
General Conditional & \texttt{If}(cond) & $1$ & \ref{sec:conditionals}\\ \hline
Protected For loop & \texttt{PFor}(init,cond,incr) & $0$ & \ref{sec:loops}\\ \hline
General For loop & \texttt{For}(init,cond,incr) & $\lceil\lg{l}\rceil$ & \ref{sec:loops}\\ \hline
General While loop & \texttt{While}(cond) & $\lceil\lg{l}\rceil$ & \ref{sec:loops}\\ \hline
Function call & \texttt{Def fxnName}(args) & $0$ & \ref{sec:fxns}\\ \hline
Log Block & \texttt{Log} &  & \ref{sec:log-unroll}\\ \hline
Unroll & \texttt{Unroll} &  & \ref{sec:log-unroll}\\ \hline

\end{tabular}

\end{table}

\textbf{Intermediate Level.} \label{sec:int-level}
The Eel intermediate language breaks down the high level control logic into jumps and labels. Jumps and labels are separated into two categories: \emph{protected} jumps and \emph{general} jumps. Protected jumps (\texttt{PGoto}, \texttt{PGotoIf}, \texttt{PGotoIfN}) are fully reversible and require no additional space in log stack. A protected conditional jump takes in a forward condition and a backward condition. It uses the assumption that the forward condition will always evaluate the same in the forward direction as the backwards condition in the reverse direction. General jumps (\texttt{Goto}, \texttt{GotoIf}, \texttt{GotoIfN}) do not require this assumption and log a bit in order to reverse. Both protected jumps and general jumps must be paired with a corresponding destination protected label or general label. Jumps and labels have a $1:1$ correspondence.

One strength of the intermediate language lies in the flexibility and variety of the jump operations. Common control logic operators of the high level can be broken down to a simple combination of protected and general jumps. This also allows new operators for the high level to be easily defined in the intermediate language without needing to touch the low level assembly-like code. Figure \ref{fig:intgrammar} shows the grammar of the intermediate language.

\begin{figure}[!htb] \label{fig:intgrammar}
\centering

\begin{grammar}

<program> ::= <b> \hfill block

<b> ::= <s>* \hfill statement sequence

<s> ::= $x$ ${\otimes}{=}$ <e> | $x =$ <e> \hfill assignment
\alt `PGoto'( $l$ )  \hfill protected jump
\alt `PGotoIf'( <e>, <e>, $l$ )
\alt `PGotoIfN'( <e>, <e>, $l$ )
\alt `PLabel'( $l$ )
\alt `Goto'( $l$ )  \hfill general jump
\alt `PGotoIf'( <e>, $l$ )
\alt `PGotoIfN'( <e>, $l$ )
\alt `Label'( $l$ )
\alt `Def' $q(x, \ldots, x)$ \hfill function definition
\alt `Call' $q(x, \ldots, x)$ \hfill function call
\alt `Log': <b> \hfill log block
\alt `Unroll' \hfill unroll
\alt `LogPush'($x$) \hfill log stack modification

<e> ::= $c$ | $x$ | <e> $\odot$ <e> \hfill expression

<$\otimes$> ::= $+$ | $-$ | $*$ \hfill operators

<$\odot$> ::= $\otimes$ | / | $\leq$ | $\geq$ | $\neq$ | $==$
\end{grammar}
\caption{Eel intermediate level grammar, where
  $x \in$ Vars, $l \in$ LabelIds, $q \in$ FxnIds, $c \in$ IntConsts}
\end{figure}

\textbf{Low Level.}\label{sec:low-level}
The low level language consists of basic assembly-level instructions that are assumed to be built into a reversible machine. Since Eel is designed for a partial reversibility model, a number of irreversible operations are also supported. Table \ref{tab:low-summary} lists the operations available at the low level. Jump operations are completely reversible and require every Goto instruction to be paired with the corresponding Comefrom instruction (\texttt{GOTOIFN} with \texttt{CMFRMIFN}). The comefrom statement is necessary in instructing the machine on bookkeeping of the program counter during jumps. 

The low level also introduces various ``special" memory locations that are reserved for specific uses. These are the program counter (\texttt{pc}), log pointer (\texttt{lp}), and stack pointer (\texttt{sp}).

\begin{table}[!htb]
\caption{Summary of low level operations.}
\label{tab:low-summary}
\centering
\begin{tabular}{|l|l|}
\columnheading{Operation} &  \columnheading{Description}  \\ \hline
\rowheading{Reversible Operations}  \\ \hline
\texttt{ADD}$(a,b)$ & $a += b$ \\ \hline
\texttt{SUB}$(a,b)$ & $a -= b$ \\ \hline
\texttt{MULT}$(a,b)$ & $a *= b$ \\ \hline
\texttt{NEG}$(a)$ & $a *= -1$ \\ \hline
\texttt{SWAP}$(a,b)$ & values $a$ and $b$ swap \\ \hline
\texttt{LPUSH}$(x)$ & push $x$ to log stack \\ \hline
\texttt{LPOP}$(x)$ & pop $x$ from log stack \\ \hline
\texttt{PUSH}$(x)$ & push $x$ to stack \\ \hline
\texttt{POP}$(x)$ & pop $x$ from stack \\ \hline
\end{tabular}
\begin{tabular}{|l|l|}
\rowheading{Irreversible Operations} \\ \hline
\texttt{MOVE}$(a,b)$ & $a = b$ \\ \hline
\texttt{AND}$(a,b)$ & $a = a\wedge b$ \\ \hline
\texttt{OR}$(a,b)$ & $a = a\vee b$ \\ \hline
\rowheading{Jump Operations} \\ \hline
\texttt{GOTO}$(l)$ & jump to $l$ \\ \hline
\texttt{GOTOIF}$(b,l)$ & jump to $l$ if $b$ \\ \hline
\texttt{GOTOIFN}$(b,l)$ & jump to $l$ if not $b$ \\ \hline
\texttt{CMFRM}$(l)$ & comefrom $l$ \\ \hline
\texttt{CMFRMIF}$(b,l)$ & comefrom $l$ if $b$ \\ \hline
\texttt{CMFRMIFN}$(b,l)$ & comefrom $l$ if not $b$ \\ \hline

\end{tabular}
\end{table}

\section{Correct Program Conventions} \label{sec:convention}

An Eel program is a code block of a sequence of statements. Statements consist of various operations and control logic which themselves can contain code blocks nested within. We model the statement execution flow of a code block as a series of forward blocks, log blocks, and unroll statements. Unroll statements trigger the reverse execution of all ``un-reversed" log blocks in the code block executed prior to the statement. If there are no pending log blocks, the Unroll statement is skipped. Every log block must be unrolled before the end of the block (synonymous to putting an unroll statement at the end of every block).

Call the set of all forward blocks, log blocks, and unrolls in a code block, $\mathcal{B}$. Let $\mathcal{B} = \mathcal{R} \: \cup \: \mathcal{F} \: \cup \: \mathcal{U}$ be the union of three distinct sets $r \in \mathcal{R}$ of log blocks, $f \in \mathcal{F}$ of forward blocks, and $u \in \mathcal{U}$ of unrolls. Every element $r$ has an element in $\mathcal{U}$ corresponding to the unroll that triggers the reverse execution of $r$, notated by $u_r$. Note that a single $u$ can satisfy the reverse execution of many $r$. The set $\mathcal{B}$ has a strict universal ordering where for all $b_i, b_j \in \mathcal{B}$, $b_i \prec b_j$ if $b_i$ occurs first in the Eel program.

Every block $b$ can be modeled as taking an input set of variables $V(b)$, executing block code, and returning the same set of variables with potentially modified values. The input and output \emph{values} of the variables are denoted $\mathcal{V}_{in}(b)$ and $\mathcal{V}_{out}(b)$. We also care about the subset of these variables that were modified, denoted by $V_{mod}(b)$. For guaranteed correct reversal of log block $r$, we desire that $\mathcal{V}_{out}(r) = \mathcal{V}_{in}(u_r)$. To receive this property, all forward blocks between $r$ and $u_r$ must not irreversibly modify any of the variables $V(r) = V(u_r)$.

\[\forall r \: \forall f \;\;\;  (\: r \prec f \prec u_r \:) \;\; \to \;\; \big( \: V(r) \cup V_{mod}(f) \: = \: \emptyset \: \big) \]

In addition to the variable modification among blocks, for a log block to be correctly reversed, control logic \emph{within} the block must be correct. This means that the requirements for all protected conditionals and for loops are satisfied. In protected conditionals, the variables in the condition cannot be modified within the conditional. In protected for loops, the variable controlling the loop cannot be modified within the loop.

It is possible for users to purposefully break these rules and still create a program that compiles and executes as they wanted. However, this requires careful variable bookkeeping and falls outside the intended use cases of the language.

\section{Control Logic Operators} \label{sec:control-logic}
Eel supports conditionals, loops, and function calls in the high level. These control logic operators are handled reversibly using the log stack. Since these operators are largely broken down in the intermediate level, we start by examining the reversibility of the jump operations. High level control operators are then built directly from intermediate jump operations, avoiding the low level. Examples are given using a log block followed by an unroll, but in general, the unroll statement need not directly follow the log block. The compilation of control logic outside of a log block is not shown here since it does not use the log stack and is compiled standardly. We note that the incorrect use of control logic and log blocks can result in an incorrect reversal and we examine this issue in the Correct Program Conventions section.

\subsection{Jumps}\label{sec:jumps}
The jump operations of the intermediate level are the building blocks for all of the high level control logic operators. The jump operations are divided into two main classes, protected jumps (fully reversible) and general jumps (require 1 logged bit). Because of their reversibility assumptions, these two classes are semantically different and are compiled differently.

Jumps are paired with labels of the same class (protected or general). In our design, we require a one-to-one pairing of jumps to labels. It is possible to support a many-to-one matching of jumps to labels, but additional information is required to be logged for reversal. Both classes support conditional jumps which use the suffixes \texttt{If} and \texttt{IfN} corresponding to jumping if the condition is non-zero or zero respectively.

In the low level, jumps can be performed by a reversible update to the program counter (\texttt{pc}). However, by allowing changes to the program counter, we can no longer assume every line was reached from the previous line by an increment to the \texttt{pc}. This creates an irreversible situation. To deal with this, every jump instruction is paired with a comefrom instruction. The Comefrom statement is used to properly handle the manipulation of the \texttt{pc}. Since the jump requires the manipulation of the \texttt{pc}, one might imagine this value being swapped or copied and manipulated. The comefrom statement performs the necessary cleaning of that value. This is necessary within the computer but not exposed at the assembly level, which is why the Comefrom simply appears to be a label or no-op.

\textbf{Protected jumps.} \label{sec:protected-jumps}
Protected jumps are fully reversible and do not use any space in the log stack. A \emph{protected} jump contains a ``backward" condition which can be evaluated in the reverse direction to indicate whether the jump was executed in the forward direction. 

Consider the protected conditional jump (\texttt{PGotoIf}). It takes the form:\\
\texttt{PGotoIf(fwdcond, bwdcond, label)}. In the forward direction, if the forward condition is true, jump to \texttt{label}. Upon reaching the label location when reversing, if the backward condition is true, jump to original jump start location. This gives the requirement that the backward condition evaluates to true if and only if the forward condition evaluated to true for the proper code to be reversed. With this assumption, we can evaluate the backward condition to determine if the jump was executed in the forward direction without additional information stored in the log stack. 

\textbf{General jumps.} \label{sec:general-jumps}
General jumps are used when the condition evaluated to decide the execution of the jump in the forward direction is not preserved and thus cannot be re-evaluated in the backward direction. In this case, we log a bit of information to the log stack to represent whether or not the jump was executed. A general jump takes the form: \texttt{GotoIf(cond, label)} where the jump to \texttt{label} is executed if \texttt{cond} is true.

In the forward direction, every time a label is reached, it was the result of either (1) increment from the line above or (2) the execution of a jump. In case (1), a 0-bit is logged, and in case (2), a 1-bit is logged. Therefore in the reverse direction, whenever a label is reached, the top bit of the log stack indicates whether to reverse the jump.

\subsection{Conditional statements}\label{sec:conditionals}
Eel high level distinguishes between two types of conditional statements, \emph{protected} conditionals and \emph{general} conditionals. In a protected conditional, the condition variables are not modified within the conditional statement.

Protected conditionals are implemented reversibly using protected jumps. If the condition variables are not modified within the conditional statement, the condition can be reevaluated after the execution of the conditional to see if the statement was executed. Thus, the condition can be used as both the forward condition and backward condition of the intermediate level protected jump. Note that this is a stronger assumption than the protected jump in the intermediate language which separates the forward and backward conditions. Figure~\ref{fig:protected-conditional} shows an example of an unsatisfied protected conditional.

The implementation of general conditionals is analogous to protected conditionals. Because the condition is subject to change in the conditional statement, the value of the condition is logged upon forward execution. This logged value is used in the backward direction to determine if the conditional statement was executed. 
\begin{figure}[!htb]
\begin{lstlisting}[language=Eel]
`High Level - Unsatisfied Protected Conditional'
Log:
  x = 1
  PIf(x):
    x -= 1
    [logic block]
Unroll
\end{lstlisting}
\caption{Example of an unsatisfied protected conditional. When reversing the condition will be \texttt{x = 0} regardless of whether the conditional statement was executed.}
\label{fig:protected-conditional}
\end{figure}

\subsection{For and While loops}\label{sec:loops}
Eel high level distinguishes between two types of for loops, \emph{protected} for loops and \emph{general} for loops. Protected for loops use no space in the log stack. Figure~\ref{fig:protected-for} shows the compilation of a protected for loop. General for loops require the number of loop iterations $l$ to be logged using $\lg{l}$ bits in the log stack.

A protected for loop takes the form: \texttt{PFor(init(x), cond(x), incr(x))}. An initial value \texttt{init(x)}, a terminating expression \texttt{cond(x)}, and a reversible incrementation function \texttt{incr(x)}. A protected for loop requires (1) the incrementation function \texttt{incr(x)} is the only modifier to \texttt{x} in the loop, and (2) the termination condition \texttt{cond(x)} is determined only by \texttt{x} and no other modified variables in the loop. With these assumptions, a protected for loop can be implemented fully reversibly. The protected for loop can be undone by reversing the incrementation function and unrolling each loop until \texttt{x} matches the initialization value. Protected jumps are used to implement the protected for loop with no space in log stack.

\begin{figure}[!htb]
\begin{lstlisting}[language=Eel]
`Protected For Loop'
`High Level'
Log:
  PFor(init(x), cond(x), incr(x)):
      [loop logic block]
    [end logic block]
Unroll
\end{lstlisting}
\begin{lstlisting}[language=Eel]
`Intermediate Level'
Log:
  init(x)
  PLabel(start-label)   //checks if x == init
  PGotoIfEq(cond(x), cond(x), end-label) //ends if cond(x)
  [loop logic block]
  incr(x)               //increments x
  PGotoIfNeq(x != init, x != init, start-label) //loops
  PLabel(end-label)
  [end logic block]
Unroll
\end{lstlisting}
\caption{The high to intermediate level compilation of a protected For loop.}
\label{fig:protected-for}
\end{figure}

A general for loop is of the form: \texttt{For(init(), cond(), incr())}. The general for loop keeps track of the number of loop iterations $l$ in the forward direction. It does not rely on the initialization variable being protected, only that the loop terminates. However, if we use general jumps, a bit of information is stored per loop and $l$ space in the log stack is required. Instead, we maintain and store a separate loop counter in the log stack using $\lg{l}$ bits. Protected jumps are then used with the general for loop condition in the forward direction and decrementation of the loop counter in the backward direction. Figure~\ref{fig:general-for} shows the compilation of a general for loop. General while loops are handled in the same way as general for loops. The initialization variable and incrementation function are disregarded.

\begin{figure}[!htb]
\begin{lstlisting}[language=Eel]
`General For Loop'
`High Level'
Log:
  For(init(), cond(), incr()):
      [loop logic block]
    [end logic block]
Unroll
\end{lstlisting}
\begin{lstlisting}[language=Eel]
`Intermediate Level'
Log:
  init()
  l = 0
  PLabel(start-label)   
  PGotoIfEq(cond(), cond(), end-label) //ends loop if cond()
  [loop logic block]
  incr()                   //incrementation function
  l += 1                   //increment loop counter
  PGotoIfEq(l > 0, l > 0, start-label) //restarts loop
  PLabel(end-label)
  LPush(l)                 //push loop counter to log
  [end logic block]
Unroll
\end{lstlisting}
\caption{The high to intermediate level compilation of a general For loop. The total number of loop iterations are counted and logged. Medium minus importance.}
\label{fig:general-for}
\end{figure}

\subsection{Function calls}\label{sec:fxns}
Reversible function calls are handled in a similar manner to normal ones. The function arguments and return pointer are pushed to the regular stack. The arguments are passed by reference, so changes to a variable effect it outside the scope of the function unless a local copy is made. Different from normal functions, for every reversible function in the high level, two versions of the function are created in the low level. One is the regular function used in the forward direction, while the other is the unrolled version used in the backward direction to reverse. Since the locations of these functions are known, protected jumps can be used to enter and exit. Thus, functions require no additional space in the log stack than what is needed for the function logic itself. Eel functions use a pointer passing parameter model taking in and modifying parameter memory locations. Figure~\ref{fig:fxn-call} shows the compilation of a function call.

\begin{figure}[!htb]
\centering

\begin{minipage}{.5\textwidth}
\begin{lstlisting}[language=Eel]
`Function Call'
`High Level'
Def FXN(x):
    [fxn logic block]

Log:
    [logic block 1]
    FXN(x)
    [logic block 2]
Unroll
\end{lstlisting}
\end{minipage}%
\begin{minipage}{.5\textwidth}
\begin{lstlisting}[language=Eel]
`Intermediate Level'
Def FXN(x):
    [fxn logic block]

Log:
    [logic block 1]
    Call FXN(x)
    [logic block 2]
Unroll
\end{lstlisting}
\end{minipage}

\begin{lstlisting}[language=Eel]
`Low Level'
//Def FXN(x):               //FXN-start
CMFRM(mem[sp-1])            //where fxn was called from
ADD(x, mem[sp-2])           //pulls input from stack
[fxn logic block]
GOTO(mem[sp-1])             //returns to program
                            //FXN-end
//Def RFXN(x):              //RFXN-start
CMFRM(mem[sp-1])            //where fxn was called from
ADD(x, mem[sp-2])           //pulls input from stack
[reverse fxn logic block]
GOTO(mem[sp-1])             //returns to program
                            //RFXN-end
[logic block 1]
PUSH(x)
PUSH(A)
GOTO(FXN-start)             //jump to fxn
CMFRM(FXN-end)              //A
POP(A)
POP(x)
[logic block 2]
[reverse of logic block 2]  //Unroll starts
PUSH(x)
PUSH(B)
GOTO(RFXN-start)            //jump to reverse fxn
CMFRM(RFXN-end)             //B
POP(B)
POP(x)
[reverse of logic block 1]
\end{lstlisting}

\caption{The full compilation of a function call. The low level shows two versions of the function for the forward direction and the backward direction.}
\label{fig:fxn-call}
\end{figure}

\section{Energy Simulation}\label{sec:energy}
Since we can't actually run our code on a semi-reversible computer, we add additional annotation to estimate the energy cost of our programs. We find this useful in two directions. First, comparing our results against theoretical predictions of the energy cost and scaling of algorithms allows us to check for inefficiencies in the compiler. Second, if our code only uses well examined transformation we can use an implementation of an algorithm as a check against the analysis of its time, space, and energy complexity.

The energy costs for an operation are defined by the change in entropy or information across the inputs to the outputs. In particular, we follow the model used in \cite{us} where one calculates $\log \left(\frac{I}{O} \right)$ Where $I$ is the size of the input space of the function and $O$ is the size of the output space of the function. This means the energy cost only depends on the instructions being called, not on the values being passed into that function. For example, this would mean an irreversible AND of two bits would always be charged $1$ unit of energy, even though an output of $0$ would tell us that both inputs had to be $0$. The appropriateness of this model either in an exact, or average case setting will depend on details of the computer architecture.

In high level programming languages, energy costs are hard to calculate since they are masked by high level control logic and complex expressions. One of the reasons Eel is designed to have multiple levels of compilation is to reveal these energy costs in the lower levels. The simplest way to calculate energy costs is in the low level language. Here the input and output spaces are small and the energy cost can be calculated on a line by line basis. Each instruction modifies only one input and since we have a restricted instruction set, each instruction's energy cost is individually evaluated. At the low level, instructions are batched into two different energy costs, 0 and $w$, where $w$ is the word size.

After calculating energy cost per line at the level, the compiler can backtrack to the intermediate and high level language and annotate each line with the costs incurred by the corresponding generated low/intermediate lines. The simulation takes the same time and space requirements of running the actual program. The annotation takes the form, \texttt{(E, L)}, representing the energy cost and space in log stack cost respectively. Logic in log blocks will incur no energy cost and instead may incur log stack cost. Conversely, logic not in log blocks will not incur any log stack cost, but may incur energy cost. 

\section{Conclusion and Future Work}

Eel is a new reversible programming language that supports a partially reversible model. The key contributions of this project are as follows:
\begin{enumerate}
    \item Development of Eel (language + compiler) and description of three language levels.
    \item Introduction of the log stack as a way to make design decisions between energy cost and space cost.
    \item Introduction of new high-level control logic and compilation techniques for protected conditional, general conditional, protected loops, and general loops.
    \item Development of an interpreter for energy simulation and annotation.
\end{enumerate}

Eel is intended to be a prototype for what partially-reversible languages may look like in the future, and to serve as a platform for the development of partially-reversible algorithms. A programming language allows us to be precise about the computations being done and serves as a platform to help verify theoretical results about partial reversibility. Because many usual programming assumptions do not hold in this model, working with Eel can help build new intuition. With the goal of algorithm development in mind, Eel has included annotation of estimated time, space, and energy costs of programs.
Through the development of the Eel language and compiler, we have built a strong foundation upon which future research and development of reversible algorithms can be conducted.
Once a few more important features are added, the ability to actually run algorithms and count the resource usage of a program will give a powerful tool for checking algorithmic results.

Several further features are necessary to achieve these goals of being a tool
for algorithmic design and a prototype language for a future computing
environment.
First, the implementation of standard data structures are necessary for many algorithms. Many of the results for efficient data structures \cite{us} in the partially reversible model are themselves not obvious and their implementation would also be a good confirmation of those results. Second, we would like to implement some of the memory management and garbage collection algorithms which have been developed. Third, only a simple version of log and unroll was implemented, which does not contain as much expressive power as we might want. Currently, the language only allows unrolling log blocks in order, but especially in data structures, we would like to be able to unroll code in dynamic orders. This extension could be implemented with multiple log stacks, or a more complicated data structure underlying the log and unroll system. Fourth, some of the transformations performed by the compiler lack optimization, and thus may make an algorithm seem less efficient than anticipated.
A final practical direction is to consider hybrid programming models which mix standard irreversible computation with reversible core subroutines, for use in a future hybrid architecture combining traditional CPUs with a reversible accelerator or co-processor.

\textbf{Acknowledgements.}
We thank Geronimo Mirano for useful discussion in differentiating and developing our language levels. We also thank Maria L. Messick and Licheng Rao for help in early programming of the Eel compiler.

\bibliographystyle{plain}
\bibliography{reversible}

\end{document}